\documentclass[twocolumn,showpacs,preprintnumbers,prb]{revtex4}

\usepackage{graphicx}
\usepackage{dcolumn}
\usepackage{bm}

\begin{document}

\title{Pressure suppression of spin-density-wave gap in the 
optical conductivity of SrFe$_2$As$_2$}

\author{H.~Okamura}\altaffiliation[Electronic address: ]{okamura@kobe-u.ac.jp}
\author{K. Shoji}
\author{K. Miyata}
\author{H. Sugawara}
\affiliation{Department of Physics, Graduate School of Science, 
Kobe University, Kobe 657-8501, Japan} 

\author{T. Moriwaki}
\author{Y. Ikemoto}
\affiliation{Japan Synchrotron Radiation Research Institute and SPring-8, 
Sayo 679-5198, Japan} 

\begin{abstract}
Optical reflectance $R(\omega)$ of a pressure-induced superconductor 
SrFe$_2$As$_2$ has been measured under external pressure to 6~GPa 
and at temperatures to 8~K.   Optical conductivity $\sigma(\omega)$ 
has been derived from the measured $R(\omega)$.  
At ambient pressure, in the antiferromagnetic state below 
$T_{\rm N}$=198~K, a pronounced feature develops in $\sigma(\omega)$ 
due to the opening of a spin density wave (SDW) gap, as already 
reported in the literature.   
With increasing pressure, the SDW gap feature in $\sigma(\omega)$ 
is progressively suppressed.   At 4~GPa, where the sample is 
superconducting, the SDW gap feature in $\sigma(\omega)$ is 
strongly reduced than that at ambient pressure, but is still 
clearly observed.   
At 6~GPa, the SDW gap is completely suppressed.  
The pressure evolutions of the SDW gap magnitude and the spectral 
weight closely follow the pressure evolution of $T_{\rm N}$.  
\end{abstract}

\pacs{ }

\maketitle

Regarding the superconductivity (SC) exhibited by Fe-based 
compounds,\cite{review} 
the pressure-induced SC exhibited 
by the stoichiometric ``122'' compounds $A$Fe$_2$As$_2$ ($A$=Ba, 
Ca, Sr, Eu) have attracted much 
interest.\cite{canfield1,canfield2,alireza,kotegawa1,isikawa,
canfield3,kotegawa2,matubayasi,igawa,yamazaki,duncan,kurita}    
They exhibit SC 
with external pressure only, although many other 
families of Fe based superconductors require some form 
of chemical doping or deviation from stoichiometry 
to exhibit SC.\cite{review}   
In particular, SrFe$_2$As$_2$ (Sr122) and BaFe$_2$As$_2$ 
(Ba122) show pressure-induced SC at relatively high 
temperatures of 
$T_c$=34~K\cite{kotegawa1,canfield3,kotegawa2,matubayasi} 
and 28~K,\cite{alireza,canfield3,matubayasi,yamazaki}, 
respectively.  
Interestingly, the 122 compounds also show SC with chemical 
doping, as shown by (Sr, K)Fe$_2$As$_2$,\cite{chu} 
Sr(Fe,Co)$_2$As$_2$,\cite{Fe-dope} and 
SrFe$_2$(As, P)$_2$.\cite{As-dope}

At ambient pressure, the stoichiometric 122 compounds 
are antiferromagnetic (AFM), poor metals.  For Sr122 and 
Ba122, the AFM 
transition temperature ($T_{\rm N}$) is 198~K and 136~K, 
respectively.   At $T_{\rm N}$, the resistivity [$\rho(T)$] shows 
a kink and rapid decrease with cooling below $T_{\rm N}$.   
With increasing pressure, $T_{\rm N}$ is gradually lowered, and 
the kink in $\rho(T)$ becomes broadened.  Above a critical 
pressure, the SC appears.    Similar suppression 
of AFM and emergence of SC have also been observed with 
chemical doping.\cite{chu,Fe-dope,As-dope}   
In this regard, both methods have provided a lot of 
insight, but one advantage of the pressure technique 
compared with chemical doping is that the former does 
not cause disorder in the crystal lattice.

However, there is an additional complication regarding the 
pressure-induced SC in the 122 compounds.  
Namely, it has been shown that 
the pressure evolutions of the AFM and SC states strongly 
depend on the hydrostaticity.\cite{kotegawa2,yamazaki,duncan}   
Here, hydrostaticity refers to how isotropic the 
pressure acting on the sample is.  
The hydrostaticity in a high pressure experiment depends 
on the types of pressure cell 
and pressure transmitting medium used.   
Figure~1 summarizes the results of high pressure 
studies on Sr122.\cite{alireza,kotegawa1,canfield3,
kotegawa2,matubayasi}  
%
\begin{figure}
\begin{center}
\includegraphics[width=0.48\textwidth]{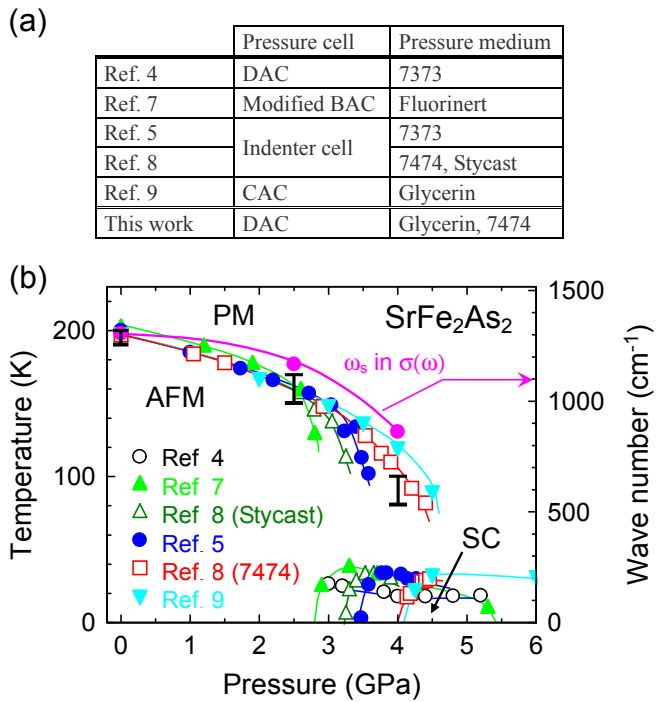} 
\caption{
(Color online) 
Summary of reported high pressure studies on SrFe$_2$As$_2$.  
(a) List of the pressure cells and pressure 
transmitting media used in previous studies.  The following 
abbreviations have been used. DAC: diamond anvil cell, 
BAC: Bridgman anvil cell, CAC: cubic anvil cell, 
7373 and 7474: Daphne Oil 7373 and 7474, respectively. 
(b) Temperature-pressure phase diagram of Sr122 suggested 
by the previous works listed in (a).  The following 
abbreviations have been used: PM: paramagnetic, 
AFM: antiferromagnetic, SC: superconductivity.  
In Ref.~4, the SC transition temperature was determined 
by magnetization, while in the other works it was 
determined by resistivity.  
The three vertical bars indicate the temperature interval 
where large decrease of optical conductivity [$\sigma(\omega)$] 
was observed in this work.   The shoulder frequency ($\omega_s$) 
in $\sigma(\omega)$ discussed in the text is also plotted.  
}
\end{center}
\end{figure}
As discussed in detail later, these results show that 
the AFM state is suppressed and SC appears at lower 
pressure when the applied pressure is less hydrostatic 
(more uniaxial).  This may indicate that a uniaxial 
stress promotes SC in the 122 compounds.

To probe the microscopic electronic structures of 
the 122 compounds, various spectroscopic techniques 
have been used.\cite{review}  
In particular, the optical conductivity [$\sigma(\omega)$] 
technique has provided much 
information.\cite{wang2,dressel2010a,hancock,leonetta,
wang1,dressel2010b,dressel2010c,degiorgi,nakajima,basov2,basov3}
As for the stoichiometric 122 compounds, a clear 
depletion of $\sigma(\omega)$ below $\sim$ 1200~cm$^{-1}$ 
was observed below $T_{\rm N}$.\cite{wang2,dressel2010a,hancock}   
This was attributed to a gapping of the Fermi surface at 
certain portions of the Brillouin Zone, which was caused 
by spin-density-wave (SDW) formation.   
At low temperatures, a narrow Drude peak due to free 
carriers was observed even after the SDW gap was formed 
in $\sigma(\omega)$.    With chemical doping into the 
122 compounds, the SDW gap in $\sigma(\omega)$ was 
progressively suppressed, and in the SC phase, a SC gap 
was clearly observed in 
$\sigma(\omega)$.\cite{wang1,dressel2010b,
dressel2010c,degiorgi,nakajima,basov2,basov3}

In this work, $\sigma(\omega)$ of Sr122 has been measured 
under external pressure ($P$) to 6~GPa and at temperature 
($T$) to 8~K in order to study the pressure evolution of 
microscopic electronic structures.   
The advantage of $\sigma(\omega)$ technique is fully 
taken here since a pressure study is technically difficult 
for other spectroscopic techniques such as photoemission 
and tunneling.    
With increasing pressure, the SDW 
gap in $\sigma(\omega)$, observed at ambient pressure as 
mentioned above, was progressively suppressed.   At 4~GPa, 
where SC is expected to appear, the SDW gap was much less 
pronounced than that at ambient pressure, but was still 
clearly observed.  At 6~GPa, where the SC was expected 
to be most stable, the SDW gap was no longer observed.

Single crystals of Sr122 were grown by a self-flux method.  
Their quality was similar to that of the samples used in 
Refs.~5 and 8, which showed SC with $T_c$=34~K at high 
pressures.     
A cleaved surface was obtained shortly before each measurement, 
which contained the plane perpendicular to the $c$ axis, and 
used without mechanical polishing.    
No polarization resolved measurement was made in this work.  
The reflectance of a sample without external pressure 
[$R_0(\omega)$] was measured in vacuum with a near-normal 
incidence over a wide frequency (photon energy) range 
between 80 and 240000~cm$^{-1}$ (10~meV and 30~eV).\cite{mybook}   
Below 20000~cm$^{-1}$, a gold film deposited {\it in situ} 
onto the sample itself was used as the reference of 
reflectance.\cite{timusk}    $\sigma(\omega)$ was derived 
from the measured $R_0(\omega)$ using the Kramres-Kronig 
(KK) analysis.\cite{dressel,wooten}  
The reflectance of a sample at high pressure 
[$R_{\rm d}(\omega)$] was measured using a diamond anvil 
cell (DAC).  
A cleaved surface of the sample was closely attached on 
the culet face of a diamond anvil.  The diameter of the 
culet face was 800~$\mu$m, and the sample dimensions were 
typically 200 $\times$ 200 $\times$ 30~$\mu$m$^3$.  
The diamond was of type IIa 
with very low density of impurities.  The sample was 
sealed in the DAC with glycerin or Daphne Oil 7474 as 
pressure transmitting medium.  The pressure was monitored 
by the ruby fluorescence method.  
$R(\omega)$ of the sample was 
measured relative to a gold film, mounted near the 
sample in the DAC.  To measure $R(\omega)$ under such 
a restricted condition, synchrotron radiation was 
used as high-brilliance infrared source\cite{IRSR-review} 
at the beam line BL43IR of SPring-8.\cite{BL43IR}   
More technical details of 
the high-pressure reflectance measurement have been 
described elsewhere.\cite{airapt}   
Note that $R_{\rm d}(\omega)$ was measured at a 
sample/diamond interface, while $R_0(\omega)$ was measured 
at a sample/vacuum interface.   Due to the large refractive 
index of diamond (2.4), therefore, $R_{\rm d}(\omega)$ may 
significantly differ from $R_0(\omega)$,\cite{footnote} 
as actually observed for a few 
compoudns.\cite{ybs,CeRu4Sb12,PrRu4P12}   Hence a careful 
analysis is required for the data measured with DAC, 
as discussed later.

Figure~2 summarizes the reflectance data of Sr122 
measured at $P$=0, 2.5, 4, and 6~GPa.    
\begin{figure}[b]
\begin{center}
\includegraphics[width=0.4\textwidth]{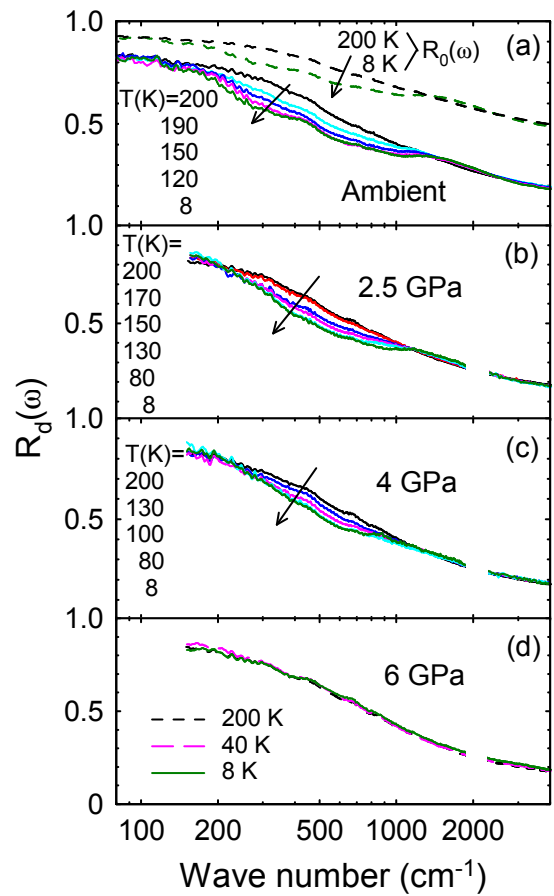} 
\end{center}
\caption{
(Color online) 
Reflectance of SrFe$_2$As$_2$ relative to diamond 
[$R_{\rm d}(\omega)$] at external pressures ($P$) 
of (a) 0, (b) 2.5~GPa, (c) 4~GPa, and (d) 6~GPa and 
at different temperatures ($T$).  In (a), 
$R_{\rm d}(\omega)$ spectra are those derived from 
the reflectance measured in vacuum 
[$R_0(w)$].\cite{footnote2}   
$R_0(\omega)$ at 295 and 8~K are also shown for 
comparison, which show that $R_{\rm d}(\omega)$ in 
the infrared is significantly lower than $R_0(\omega)$ 
due to the large refractive index of diamond.  
In (b)-(d), the spectra in the 1870-2350~cm$^{-1}$ 
range are not shown since this range could not be 
measured due to strong absorption by the diamond 
anvil.  This range was interpolated by a straight 
line for the KK analysis to obtain $\sigma(\omega)$.  
}
\end{figure}
These spectra were derived from measured data using 
the procedures previously discussed.\cite{footnote2}    
Note that the $R_{\rm d}(\omega)$ spectra, expected 
from the measured $R_0(\omega)$,\cite{footnote2} are 
indicated even for the $P$=0 case [Fig.~2(a)], to 
enable direct comparison with the high pressure cases 
[Fig.~2(b)--2(d)].  
At $P$=0, in Fig.~2(a), a strong reduction of reflectance 
below approximately 1200~cm$^{-1}$ is observed at 
$T<$ 200~K, where the transition to 
AFM state and the formation of SDW state occur.  This 
reduction of reflectance below $T_{\rm N}$ is due to a 
formation of SDW gap, 
and has been analyzed and discussed by published 
works, including similar results on Ba122.
\cite{wang2,dressel2010a,hancock}   
With increasing $P$, in Figs.~2(b)-2(d), the reduction 
of reflectance 
becomes smaller, and it is absent at $P$=6~GPa.  
These pressure evolutions indicate a suppression of 
the SDW gap, as discussed in detail 
below based on the $\sigma(\omega)$ data.  
Note that we did not observe significant difference 
between data obtained with glycerin and Daphne oil 
7474 as pressure medium.

Figure~3 shows the $\sigma(\omega)$ spectra obtained 
from the reflectance data in Fig.~2.   
\begin{figure}[b]
\begin{center}
\includegraphics[width=0.4\textwidth]{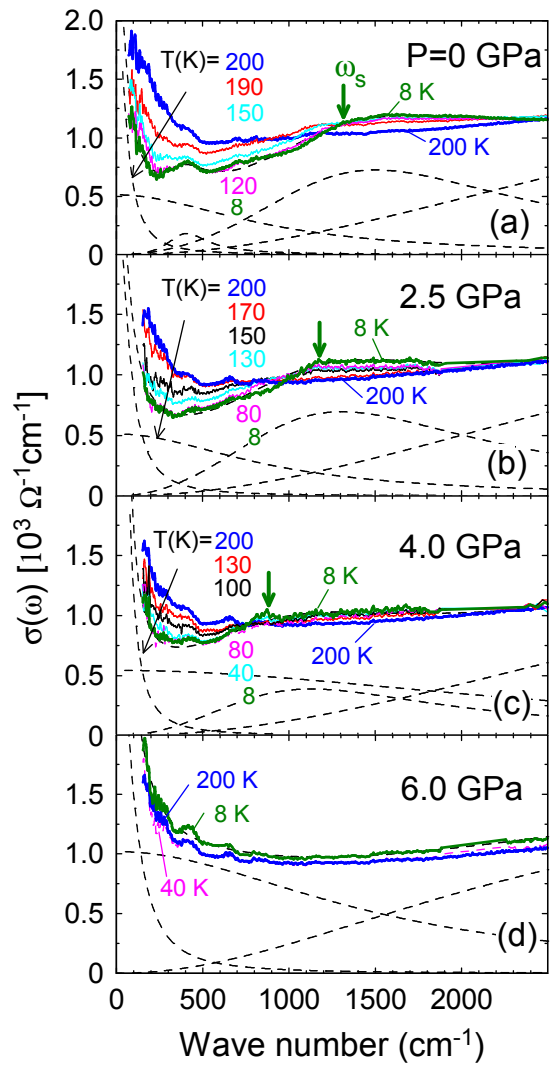} 
\end{center}
\caption{
(Color online) 
Optical conductivity [$\sigma(\omega)$] spectra of 
SrFe$_2$As$_2$ at external pressures ($P$) of (a) 0, 
(b) 2.5~GPa, (c) 4~GPa, and (d) 6~GPa and at 
different temperatures ($T$).  $\omega_s$ and the 
vertical arrows indicate the shoulder frequency.  
The broken curves show the results of Drude-Lorentz 
fitting for the 8~K data as discussed in the text.  
}
\end{figure}
Here, the $\sigma(\omega)$ spectra at $P$=0 in Fig.~3(a) were 
obtained from $R_0(\omega)$ using the conventional KK 
analysis,\cite{dressel,wooten} 
while those at high pressure in Figs.~3(b)-3(d) were 
obtained from $R_{\rm d}(\omega)$ using a modified KK analysis, 
which took into account the refractive index of 
diamond as previously discussed in detail.\cite{kk}   
At $P$=0 and $T$=200~K, in Fig.~3(a), $\sigma(\omega)$ 
is almost flat above 500~cm$^{-1}$, and rises rapidly 
with decreasing frequency.  The rising component 
is due to the Drude response of free 
carriers.\cite{dressel,wooten}   
With cooling below $T_{\rm N}$=198~K, $\sigma(\omega)$ 
below $\sim$ 1200~cm$^{-1}$ is depleted, and with 
further cooling, $\sigma(\omega)$ develops an gap-like 
structure.   However, even at 8~K, the spectral depletion 
is incomplete, and $\sigma(\omega)$ seems to contain a 
flat continuum or background.   In addition, the Drude 
response becomes narrower with cooling, and a shoulder 
appears near 1300~cm$^{-1}$ as indicated by the 
vertical arrow in Fig.~3(a).    This shoulder marks 
the characteristic frequency in $\sigma(\omega)$ below 
which the spectral depletion occurs.   
The formation of incomplete gap in $\sigma(\omega)$ 
below $T_{\rm N}$ has been observed and discussed for 
both Sr122 and Ba122 by many 
authors.\cite{wang2,dressel2010a,hancock,degiorgi,nakajima}   
Note that, as previously discussed, the Drude 
response remains even at 8~K.   From these results, 
it has been concluded that a gapping of Fermi surface 
occurs at certain portions of the Brillouin Zone due 
to the SDW formation, and that the electronic 
subsystem responsible for the narrow Drude response 
is distinct from that for the SDW gap 
opening.\cite{dressel2010a,nakajima}   
At $P$=2.5~GPa, in Fig.~3(b), the spectral depletion 
of $\sigma(\omega)$ with cooling is still seen, but a 
large decrease is observed below 170~K, rather than 
below 200~K as in the $P$=0 data.   This result well 
corresponds to the decrease of $T_{\rm N}$ from 198~K 
at $P$=0 to about 160~K at 
$P$=2.5~GPa.\cite{kotegawa1,kotegawa2}  
The spectral range of depletion marked by the position of 
the shoulder, which is about 1200~cm$^{-1}$, is also 
smaller than that at $P$=0.  These trends continue 
in the 4~GPa data.  Namely, the spectral depletion becomes 
smaller, and occurs over narrower range with a 
shoulder near 900~cm$^{-1}$.  Along these pressure 
evolutions, the narrow Drude response mentioned above 
is surviving at 4~GPa.   At 6~GPa, in Fig.~3(d), in contrast, 
$\sigma(\omega)$ does not show spectral depletion any more, 
and shows small increase with cooling.   
$\sigma(\omega)$ spectra at different pressures at 
$T$=200~K and 8~K are indicated in Fig.~4.   
\begin{figure}[b]
\begin{center}
\includegraphics[width=0.4\textwidth]{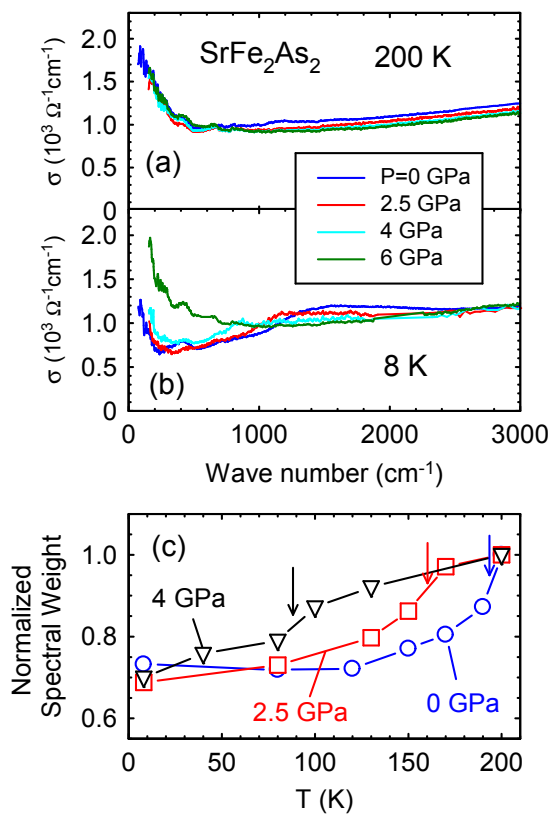} 
\end{center}
\caption{
(Color online) 
Optical conductivity [$\sigma(\omega)$] spectra of 
SrFe$_2$As$_2$ at (a) 200~K and (b) 8~K at different 
external pressures ($P$).  
(c) Normalized spectral weight of $\sigma(\omega)$ 
showing the depletion due to SDW gap formation as a 
function of temperature ($T$).   Here, the average 
of $\sigma(\omega)$ at frequencies between 
$0.4 \omega_s$ and $0.6 \omega_s$ normalized by 
$\sigma(\omega_s)$ at 200~K, where $\omega_s$ is the 
shoulder frequency in Fig.~3, is plotted versus $T$.  
The vertical arrows indicate large drops of 
$\sigma(\omega)$ due to the gap formation.  
}
\end{figure}
The shift of shoulder and the suppression 
of the SDW gap with pressure are more clearly seen.  
The shoulder frequency ($\omega_s$) is plotted as a 
function of pressure in the phase diagram of Fig.~1(a).  
It is seen that the pressure evolution of $\omega_s$ 
closely follows that of $T_{\rm N}$, i.e., the pressure 
evolution of SDW gap 
magnitude also closely follows that of $T_{\rm N}$.

According to the previous works on Sr122 (Fig.~1), 
SC appears at approximately 4~GPa when the pressure 
is applied by cubic anvil and indenter cells with 
glycerin and Daphne Oil (DO) 7474 as pressure transmitting 
media, respectively.\cite{kotegawa1,kotegawa2,matubayasi}    
SC appears at lower pressure, at $\sim$ 3~GPa, when 
DAC and Bridgman anvil cells are 
used with Fluorinert and DO 7373, respectively.
\cite{alireza,canfield3}  
   Cubic anvil and indenter cells generally produce 
more hydrostatic pressure than Bridgeman and diamond 
anvil cells if the same pressure medium is used.   
In addition, glycerin and DO 7474, which are also used 
in the present study, produce more hydrostatic pressure 
than DO 7373 and Fluorinert if the same pressure cell 
is used.  
(Hydrostaticity of these pressure transmitting 
media in DAC have been studied in detail.
\cite{osakabe,7474,tateiwa,klotz1,klotz2,footnote3})  
Based on these considerations and the results in 
Fig.~1, it is reasonable to assume that the 
sample be in the SC state at 4 and 6~GPa in this study, 
although no resistivity measurement has been made.    
For chemically doped 122 systems, the formation of a 
SC gap has been manifested by a strong depletion of 
$\sigma(\omega)$ below $\sim$ 200~cm$^{-1}$, and by 
an increase of reflectance toward unity.
\cite{wang1,degiorgi,nakajima,basov3}  
In the present data of $\sigma(\omega)$ and $R(\omega)$ 
at 4 and 6~GPa, however, no such features are observed, 
although the sample should be in SC state as discussed 
above.  The reason why a SC gap is not observed in 
our high pressure $\sigma(\omega)$ is unclear.   
A possible reason is the limited spectral range of 
the present study: the data in this work could be 
measured only above 150~cm$^{-1}$ due to 
the technical limitations associated with DAC.   
In previous works with chemically doped samples, 
measurements were made to much lower frequencies, 
generally down to 50~cm$^{-1}$ or lower.
\cite{wang1,degiorgi,nakajima,basov3}

In the previous optical studies of 122 compounds, 
Drude-Lorentz spectral fitting\cite{dressel,wooten} was 
used to analyze the $\sigma(\omega)$ 
spectra.\cite{wang2,dressel2010a,hancock,degiorgi,nakajima}   
Results of similar fitting to our high pressure 
data at 8~K are indicated by the broken curves in Fig.~3.  
Note that, in addition to a narrow Drude 
term for the narrow Drude response in $\sigma(\omega)$, a 
broad Drude term is used in the fitting function for 
the broad continuum in $\sigma(\omega)$, as previously 
done.   A Lorentz term is used to fit the gap feature 
including the shoulder, and another Lorentz term 
for a high frequency background.   [For $P$=0 data, 
a small Lorentz term is also used to fit the small 
peak near 400~cm$^{-1}$.]   
With increasing pressure, the spectral weight (SW) 
of the Drude terms increase, and that of the Lorentz 
term for the gap feature decreases.   This shows an 
increase of carrier density in correspondence to the 
suppression of SDW gap with pressure.   
These evolutions with pressure are qualitatively similar 
to those with increasing $T$ at $P$=0, in Fig.~3(a), and 
similar to those with increasing chemical 
doping.\cite{degiorgi,nakajima}     In the previous works, 
the carrier scattering rate ($\gamma$) and SW of the 
narrow Drude term were obtained from the 
fitting.\cite{wang2,dressel2010a,degiorgi,nakajima}  
$\gamma$ was strongly suppressed when the SDW gap was 
formed, and the effective carrier density was also 
evaluated from SW.   
We have attempted similar analysis, but have found 
large uncertainty in the fitting parameters obtained.  
This is primarily due to the limited spectral range 
of our high pressure experiment: The low frequency 
end of our study, limited by the use of DAC, was 
150~cm$^{-1}$, which was too high for a reliable fitting 
of the narrow Drude response.   
[Only the tail of the narrow Drude response is 
seen in $\sigma(\omega)$, especially at low $T$.]    
Hence, it was difficult to accurately evaluate the 
pressure evolution of $\gamma$ and SW of the 
carriers.

To analyze the evolution of SDW gap with $P$ and $T$, 
we have plotted in Fig.~4(c) the SW of $\sigma(\omega)$ 
below the shoulder.    
Here, the average of measured $\sigma(\omega)$ at frequencies 
between $0.4 \omega_s$ and $0.6 \omega_s$ normalized 
by $\sigma(\omega_s)$ at 200~K, where $\omega_s$ 
is the shoulder frequency, is plotted as a function 
of $T$.   (The results of fitting were not used 
here due to the uncertainty discussed above.)  
The choice of the $0.4 \omega - 0.6 \omega_s$ range 
is rather arbitrary, and is intended to probe a 
region in the gap but without a significant overlap 
with the narrow Drude component.  
It is seen that the spectral weight in the gap steeply 
drops with cooling below 170~K at 2.5~GPa, and below 
100~K at 4~GPa, respectively.   These $T$ ranges where 
the steep decrease of $\sigma(\omega)$ is seen are 
indicated by the vertical bars in Fig.~1(b).   
It is clear that they well follow the evolution 
of $T_{\rm N}$ with pressure.

In summary, $\sigma(\omega)$ of pressure-induced 
superconductor Sr122 was obtained under external 
pressure to 6~GPa using diamond anvil cell.   
The SDW gap formed in $\sigma(\omega)$ at ambient 
pressure was progressively suppressed with increasing 
pressure.   The SDW gap was still present at 4~GPa 
where the sample was expected to be in SC state.  
At 6~GPa, where the SC was expected to be optimum, 
the SDW gap was completely suppressed.   
The pressure evolution of the SDW gap magnitude, 
implied by a shoulder in $\sigma(\omega)$, and 
that of the spectral weight within the SDW gap, 
closely followed the evolution of $T_N$.  
In addition, the spectral evolution of the SDW gap 
with increasing pressure was qualitatively similar to 
those with increasing temperature at $P$=0, and also 
to those with increasing chemical pressure at $P$=0.   
A SC gap was not observed in our $\sigma(\omega)$ data 
even at 4 and 6~GPa where the sample should be in 
the SC state, probably due to a limited spectral range 
of our high pressure study using DAC.

\begin{acknowledgements}
H. O. acknowledges Prof. H. Kotegawa for useful 
information about high pressure studies on Sr122.     
The work at SPring-8 was made under the approval of JASRI 
(2009A0089 through 2012A0089).  Financial support from 
MEXT (''Heavy Fermion'' 21102512-A01 and Scientific 
Research C 23540409) is acknowledged.  
\end{acknowledgements}



\begin{thebibliography}{99}
\bibitem{footnote}
The normal-incidence reflectance of a sample surface 
is given by the Fresnel's formula as 
\begin{equation}
R(\omega)= \left|
\frac{n_0 - \hat{n}(\omega)}{n_0 + \hat{n}(\omega)}
\right|^2. 
\end{equation}
Here, $\hat{n}(\omega)$ is the complex refractive index of 
the sample, and $n_0$ is the refractive index of the diamond 
(2.4) in the case of $R_{\rm d}(\omega)$, and 1.0 for $R_0(\omega)$.   
Therefore, $R_{\rm d}(\omega)$ may significantly differ 
from $R_0(\omega)$.
\bibitem{footnote3}
The solidification pressures at room temperature 
have been reported as follows.  
Fluorinert : 1~GPa\cite{7474}; 
Daphne 7373: 2.2~GPa\cite{osakabe,7474}; 
Daphne 7474: 3.7~GPa\cite{7474,klotz1}; and 
glycerin: 5.5~GPa.\cite{klotz2}
\end{thebibliography}
\end{document}